\def\year{2019}\relax
\documentclass[sigconf]{acmart}
\usepackage{url}  
\usepackage{graphicx}  
\usepackage{amsmath}
\usepackage{flushend}

\usepackage[ruled,longend]{algorithm2e}

\setcopyright{acmlicensed}

\begin{document}

\title{Discovering Communities of Community Discovery}

\author{Michele Coscia}
\email{mcos@itu.dk}
\affiliation{%
  \institution{IT University of Copenhagen}
  \streetaddress{Rued Langgaards Vej 7}
}

\begin{abstract}
Discovering communities in complex networks means grouping nodes similar to each other, to uncover latent information about them. There are hundreds of different algorithms to solve the community detection task, each with its own understanding and definition of what a ``community'' is. Dozens of review works attempt to order such a diverse landscape -- classifying community discovery algorithms by the \textbf{process} they employ to detect communities, by their explicitly stated \textbf{definition} of community, or by their \textbf{performance} on a standardized task. In this paper, we  classify community discovery algorithms according to a fourth criterion: the \textbf{similarity} of their results. We create an Algorithm Similarity Network (ASN), whose nodes are the community detection approaches, connected if they return similar groupings. We then perform community detection on this network, grouping algorithms that consistently return the same partitions or overlapping coverage over a span of more than one thousand synthetic and real world networks. This paper is an attempt to create a similarity-based classification of community detection algorithms based on empirical data. It improves over the state of the art by comparing more than seventy approaches, discovering that the ASN contains well-separated groups, making it a sensible tool for practitioners, aiding their choice of algorithms fitting their analytic needs.
\end{abstract}

\copyrightyear{2019}
\acmYear{2019}
\acmConference[ASONAM '19]{International Conference on Advances in Social Networks Analysis and Mining}{August 27--30, 2019}{Vancouver, BC, Canada}
\acmBooktitle{International Conference on Advances in Social Networks Analysis and Mining (ASONAM '19), August 27--30, 2019, Vancouver, BC, Canada}
\acmPrice{15.00}
\acmDOI{10.1145/3341161.3342860}
\acmISBN{978-1-4503-6868-1/19/08}

\maketitle

\section{Introduction}
In this paper, we provide a bottom-up data-driven categorization of community detection algorithms. Community detection in complex networks is the task of finding groups of nodes that are closely related to each other. Doing so usually unveils new knowledge about how nodes connect, helping us predicting new links or some latent node characteristic.

Community discovery is probably the most prominent and studied problem in network science. This popularity implies that the number of different networks to which community discovery can be applied is vast and so is the number of its potential analytic objectives. As a result, what a community \textit{is} in a complex network can take as many different interpretations as the number of people working in the field.

Review works on the topic abound and often their reference lists contain hundreds of citations \cite{fortunato2010community}. They usually attempt a classification, grouping community detection algorithms into a manageable set of macro categories. Most of them work towards one of three  objectives. They classify community detection algorithms: by \textbf{process}, meaning they explain the inner workings of an algorithm and let the reader decide which method corresponds to their own definition of community -- e.g. \cite{fortunato2010community}; by \textbf{definition}, meaning they collect all community discovery definitions ever proposed and create an ontology of them -- e.g. \cite{coscia2011classification}; by \textbf{performance}, meaning that they put the algorithms to a standardized task and rank them according to how well they perform on that task -- e.g. \cite{hric2014community}.

This paper also attempts to classify community discovery algorithms, but uses none of these approaches. Instead, we perform a categorization by \textbf{similarity}, e.g. which algorithms, at a practical level, return almost the same communities. As in the process case, we expect the inner workings of an algorithm to make most of the difference, but we do not focus on them. As in the definition case, we aim to build an ontology, but ours is bottom-up data-driven rather than being imposed top-down. As in the performance case, we define a set of standardized tasks, but we are not interested in which method maximizes a quality function.  

Here, we are not interested in what works \textit{best} but what works \textit{similarly}. This is useful for practitioners because they might have identified an algorithm that finds the communities they are interested in, with some downsides that make its application impossible (e.g. long running times). With the map provided in this paper, a researcher can identify the set of algorithms outputting almost identical results to their favorite one, but not affected by its specific issues. Maybe they perform slightly worse, but do so at a higher time efficiency.

We do so by collecting implementations of community detection algorithms and extract communities on synthetic benchmarks and real world networks. We then calculate the pairwise similarity of the output groupings, using overlapping mutual information \cite{lancichinetti2009detecting}, \cite{mcdaid2011normalized} -- we need the overlapping variant, because it allows us to compare algorithms which allow communities to share nodes. For each network in which algorithms $a_1$ and $a_2$ ranked in the top five among the most similar outputs we increase their similarity count by one. 

Once we have an overall measure of how many times two algorithms provided similar communities, we can reconstruct an affinity graph, which we call the Algorithm Similarity Network ($ASN$). In $ASN$, each node is a community discovery method. We weigh each link according to the similarity count, as explained above. We only keep links if this count is significantly different from null expectation. Once we establish that our reconstruction of $ASN$ is resilient to noise and to our choices, we analyze it. Specifically, we want to find groups of algorithms that work similarly: we discover communities of community discovery algorithms.

There are other approaches proposing a data-driven classification of community discovery algorithms \cite{dao2018estimating, dao2018community, ghasemian2018evaluating}. This paper improves over the state of the art by: exploring more algorithms (73) over more benchmarks (960 synthetic and 819 real-world networks) than other empirical tests; exploring more algorithm types -- including overlapping and hierarchical solutions --; looking at the actual similarity of the partitions rather than the distribution of community sizes.

Note that we were only able to collect 73 out of the hundreds community discovery algorithms, because we focused on the papers which provided an easy way to recover their implementation. This paper should not be considered finished as is, but rather as a work in progress. Many prominent algorithms were excluded as it was not possible to find a working implementation -- sometimes because they are simply too old. Authors of excluded methods should be assured that we will include their algorithm in $ASN$ if they can contact us at \url{mcos@itu.dk}. The most updated version of $ASN$ will then be not in this paper, but available at \url{http://www.michelecoscia.com/?page_id=1640}.

\section{Related Work}
This paper fits into the vast literature of community discovery, specifically among those papers that try to organize it into a reduced set of categories that can be understood and used by practitioners. Community detection in complex networks is a prolific field with hundreds of different approaches and dozens of different community definitions. Such review works are a necessary element to make the field manageable. We can classify reviews into four categories, each of which focusing on a different aspect of community discovery.

The first -- most popular -- category includes works classifying algorithms by the techniques they employ to divide the graph into groups of nodes, i.e. by their \textbf{process}. Examples in this category are \cite{newman2004detecting}, \cite{porter2009communities}, \cite{fortunato2010community}, \cite{parthasarathy2011community}, \cite{fortunato2016community}, \cite{fang2019survey}; \cite{kim2015community} -- focusing on multilayer networks; and \cite{cai2016survey} -- whose attention narrows down to genetic algorithms. Here, we are agnostic about how an algorithm works, as we are focused on figuring out which algorithm returns similar partitions to which other. This is influenced by how they work, but even algorithms based on the philosophy of modularity maximization might end up in different categories.

The second category includes works classifying community discovery algorithms by the \textbf{definition} of community they are searching for in the network. Notable definition-based review works are \cite{schaeffer2007graph}, \cite{coscia2011classification}, \cite{malliaros2013clustering}, \cite{amelio2014overlapping}, and \cite{rossetti2018community}, the latter three focusing on directed, overlapping, and evolving networks. This is the closest category to ours, as we are also interested in building an ontology of community discovery algorithms. However, the works in this category employ a top-down approach. They take the stated -- theoretical -- definition of community of a paper and use it to classify it. Here, we have a data-driven approach: we classify algorithms not by their stated definition, but by their practical results.

The third category -- gaining popularity recently -- includes works classifying community discovery algorithms by giving them a specific task and ranking them in how well they \textbf{perform} in that task. Such tasks can be maximizing modularity or the normalized mutual information of the communities they recover versus some other metadata we have about the nodes. In this category, we can find papers such as \cite{danon2005comparing}, \cite{lancichinetti2009community}, \cite{orman2009comparison}, \cite{leskovec2010empirical}, \cite{hric2014community}, \cite{harenberg2014community}, \cite{yang2016comparative}; and, specifically for overlapping community discovery, \cite{xie2013overlapping}. In line with this approach, we also use standardized tests and benchmarks. However, we have no interest in which algorithm performs ``best'' -- whatever the definition of ``best'' is -- rather in what works similarly. We have a small ranking discussion, but we use it to criticize the notion of a ``best'' community discovery algorithm rather than taking the results at face value.

The final, and least explored, category is interested in classifying the community discovery algorithms by the \textbf{similarity} of their outputs \cite{dao2018estimating, dao2018community, ghasemian2018evaluating}. This is where our paper belongs. The typical paper in this category tests a handful of algorithms on a limited number of synthetic or real world networks. Here we include 73 algorithms\footnote{Links and references: \url{http://www.michelecoscia.com/?page_id=1640}.} -- which is the highest number of methods considered empirically -- over more than a thousand benchmark networks. This is not just a quantitative improvement: by having more algorithms we are also able to include a more diverse set of algorithms, with different features. This makes our results a better picture of the landscape of community detection in complex networks.

\section{Method}
The aim of this paper is to build an Algorithm Similarity Network ($ASN$), whose elements are the similarities between the outputs of community discovery algorithms. To evaluate result similarity is far from trivial, as we need to: (i) test enough scenarios to get a robust similarity measure, and (ii) being able to compare disjoint partitions to overlapping coverages -- where nodes can be part of multiple communities.

In this section we outline our methodology to build $ASN$, in three phases: (i) creating benchmark networks; (ii) evaluating the pairwise similarity of results on the benchmark networks; and (iii) extracting $ASN$'s backbone.

A note about generating the results for each algorithm. Many algorithms require parameters and do not have an explicit test for choosing the optimal ones. In those cases, we operate a grid search, selecting the combination yielding the maximum modularity. This is simpler in the case of algorithms returning disjoint partitions. For algorithms providing an overlapping coverage, there are multiple conflicting definitions of overlapping modularity. For this paper, we choose the one presented in \cite{lazar2010modularity}. 

\subsection{Benchmarks}
We have two distinct sets of benchmarks on which to test our community discovery algorithms: synthetic networks and real world networks.

\subsubsection{Synthetic Networks}
In evaluating community discovery algorithms, most researchers agree on using the LFR benchmark generator \cite{lancichinetti2008benchmark} for synthetic testing. The LFR benchmark creates networks respecting most of the properties of interest of many real world networks. We follow the literature and use the LFR benchmark. We make this choice not without criticism, which we spell out in Section \ref{sec:anal-robust}.

To generate an LFR benchmark we need to specify several parameters. Here we focus on two in particular: number of nodes $n$ and mixing parameter $\mu$ -- which is the fraction of edges that span across communities, making the task of finding communities harder. We create a grid, generating networks with $n = \{50, 60, 70, 80, 90, 100\}$ and $\mu = \{.07, .09, .11, .13, .15, .17, .19, .21\}$. The average degree ($\bar{k}$) is set to 6 for all networks, while the maximum degree ($K$) is a function of $n$. For each combination of parameters we generate ten independent benchmarks with disjoint communities and ten benchmarks with overlapping communities. In the overlapping case, the number of nodes overlapping between communities ($o_n$), as well as the number of communities to which they belong ($o_m$), are also a function of $n$.

We generate 2 (overlapping, disjoint) $\times$ 10 (independent benchmarks) $\times$ 6 (possible number of nodes) $\times$ 8 (distinct $\mu$ values) $=$ 960 benchmarks. Due to the high number of networks and to the high time complexity of some of the methods, we are unable to have larger benchmarks. The number of benchmarks is necessary to guarantee statistical power to our similarity measure.

\subsubsection{Real World Networks}
The LFR benchmarks have a single definition of community in mind. Therefore the tests are not independent, and if an algorithm follows a different community definition, it might fail in unpredictable ways, which makes our edge creating process prone to noise.

To reduce this issue, we collect a number of different real world networks. Communities in real world networks might originate from a vast and variegated set of possible processes. We assembled 819 real world networks, which were found in the Colorado Index of Complex networks\footnote{\url{https://icon.colorado.edu/}. Complete reference list: \url{http://www.michelecoscia.com/?page_id=1640}.}. We selected a high number of small networks to conform to our needs of statistical significance as described in the previous subsection.

\subsection{Evaluating Similarity}\label{sec:method-simil}
Once we run two community discovery algorithms on a network, we obtain two divisions of nodes into communities. A standard way to estimate how similar these two groupings are is to use normalized mutual information \cite{vinh2010information} (NMI). Mutual information quantifies the information obtained about one random variable through observing the other. The normalized variant, rather than returning the number of bits, is adjusted to take values between 0 (no mutual information) and 1 (perfect correlation).

The standard version of NMI is defined only for disjoint partitions, where nodes can belong to only one community. However, many of the algorithms we test are overlapping, placing nodes in multiple communities. There are several ways to extend NMI to the overlapping case  (oNMI), as described in \cite{lancichinetti2009detecting} and \cite{mcdaid2011normalized}. We use the three definitions considered in these two papers as our alternative similarity measures. These versions reduce to NMI when their input is two disjoint partitions. This allows us to compare disjoint and overlapping partitions to each other.

We label the three variants as MAX, LFK, and SUM, following the original papers. Our default choice is MAX, which normalizes the mutual information between the overlapping results $a_1$ and $a_2$ with the maximum of the entropy of $a_1$ and $a_2$. Differently from LFK, MAX is corrected by chance: unrelated vectors will have zero oNMI MAX.

How do we aggregate the similarity results across our 1,779 benchmarks? We have three options: (i) averaging them, (ii) counting the number of times two algorithms had an oNMI higher than a given threshold, and (iii) counting the number of times two algorithms were each other in the most similar algorithms in a given benchmark. We choose option (iii).

Option (i) has both theoretical and practical issues. It is not immediately clear what is the semantic of an average normalized mutual information. Moreover, we want to empathize the scenarios in which two algorithms are similar more than when they are dissimilar. There is only one way in which two results can be similar, while there are (almost) infinite ways for two results to be dissimilar. Thus similarity contains more information than dissimilarity. If we take the simple average, dissimilarity is going to drive the results.

In option (ii), NMIs will have different expected values for different networks. If we choose a threshold for all benchmarks, we will overweight some benchmarks over others. This is fixed by option (iii), which counts the cases in which both algorithms agree on the community structure in the network.

Note that both algorithms have to agree, thus this method still allows algorithms to be isolated if they are dissimilar to everything else. Suppose $a_1$ is a very peculiar algorithm. Regardless of its results, it will find $a_2$ as its most similar companion, even if the results are different. Since the results are different, $a_2$ will not have $a_1$ as one of its most similar companions. Thus there will be no edge between $a_1$ and $a_2$.

We will see in our robustness checks that the three options return comparable results, with option (iii) having the fewest theoretical and practical concerns.

\subsection{Building the Network}
The result from the previous section is a weighted network, where each edge weight is the number of benchmarks in which two algorithms were in each other most similar results. Any edge generation choice will generate a certain amount of noise. Algorithms with average results might end up as most similar to other algorithms in a benchmark just by pure chance. This means that there is uncertainity in our estimation of the edge weights -- or whether some edges should be present at all.

To alleviate the problem, we use the noise corrected (NC) backbone approach \cite{coscia2017network}. The reason to pick this approach over the alternatives lies in its design. The NC backboning procedure removes noise from edge weight estimates, under specific assumptions about the edge generation process, which fit the way we build our network. $ASN$ is a network where edge weights are counts, broadly distributed -- as we show in the Analysis section --, and are generated with an hypergeometric ``extraction without replacement'' approach, which are all assumptions of the NC backboning approach. For this reason, we apply the NC backbone to our $ASN$.

The NC backbone requires a parameter $\delta$, which controls for the statistical significance of the edges we include in the resulting network. We set the parameter to the value required to have the minimum possible number of edges, while at the same time ensuring that each node has at least one connection. In our case, we set $\delta = 19.5$, meaning that we only include edges with that particular t-score (or higher), which is roughly equivalent to say that $p < .00001$.

Again, note that we are not imposing the $ASN$ to be connected in a single component. Under these constraints, $ASN$ could be just a set of small components, each composed by a pair of connected algorithms.

\section{Analysis}

\subsection{The Algorithm Similarity Network}

\begin{figure}
\centering
\includegraphics[width=\columnwidth]{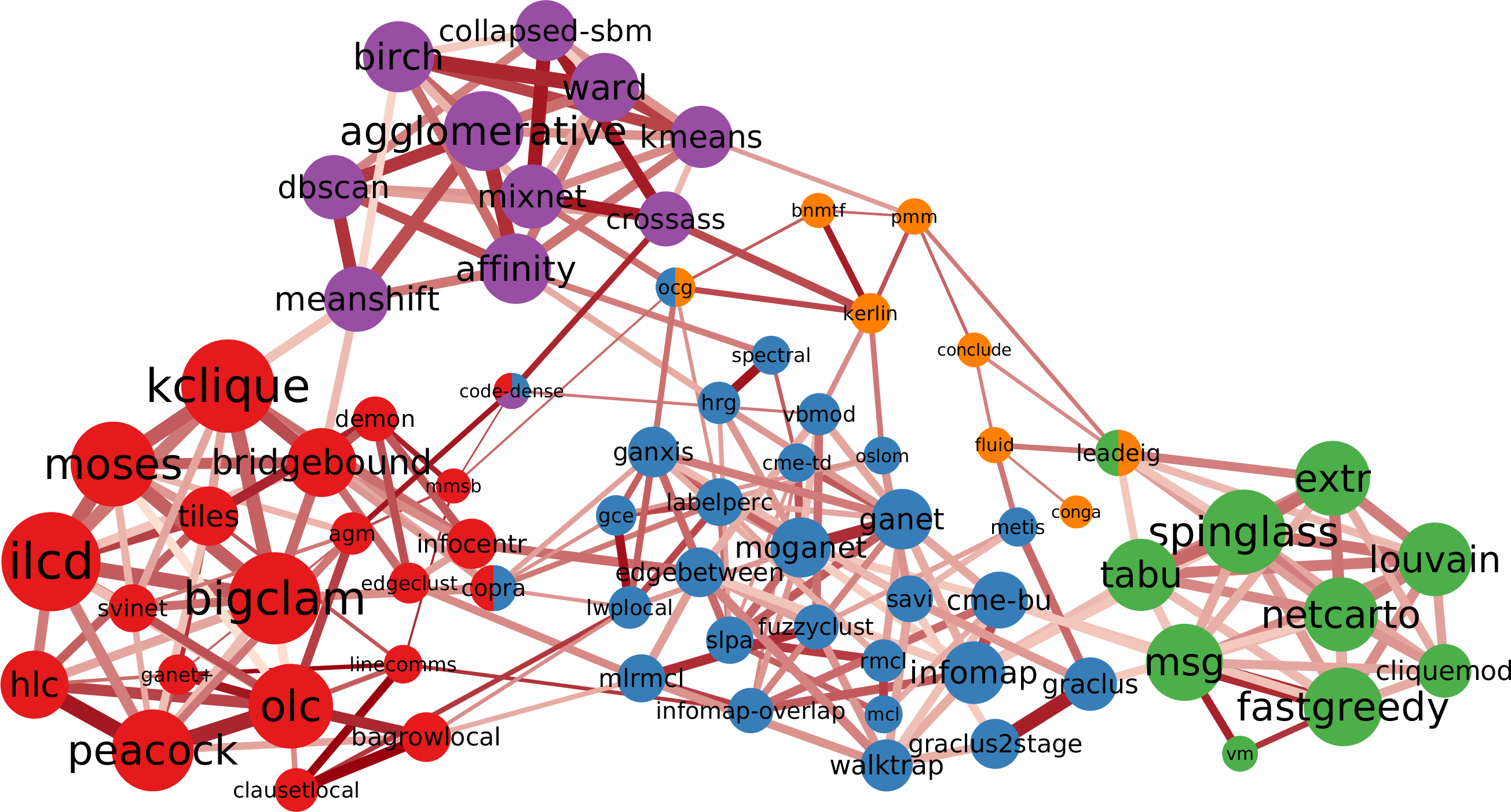}
\caption{$ASN$. Nodes are community detection algorithms. Node size: sum of total edge weights. Node color: community affiliation -- multicolored nodes belong to multiple communities. Edge width: number of times the two algorithms returned similar partitions. Only including links exceeding null expectation. Link color: significance, from dark (high) to light (low, but still significant with $p < .00001$).}
\label{fig:asn}
\end{figure}

We start by taking a look at the resulting $ASN$ network. We show a depiction of the network in Figure \ref{fig:asn} -- calculated using the oNMI MAX similarity function and setting $\delta = 19.5$ for the noise corrected backboning. The network contains all the results, both from synthetic and from real-world networks.

The first remarkable thing about $ASN$ is that it does have a community structure. The network is sparse -- by construction, this is not a result --: only 9\% of possible edges are in the network. However, and this is surprising, clustering is high -- transitivity is 0.47, or 47\% of connected node triads have all three edges necessary to close the triangle.

For these reasons, we can run a community discovery algorithm on $ASN$. We choose to run the overlapping Infomap algorithm \cite{rosvall2008maps}. The algorithm attempts to compress the information about random walks on the network using community prefix codes: good communities compress the walks better because the random walker is ``trapped'' inside them.

\begin{figure}
\centering
\includegraphics[width=.66\columnwidth]{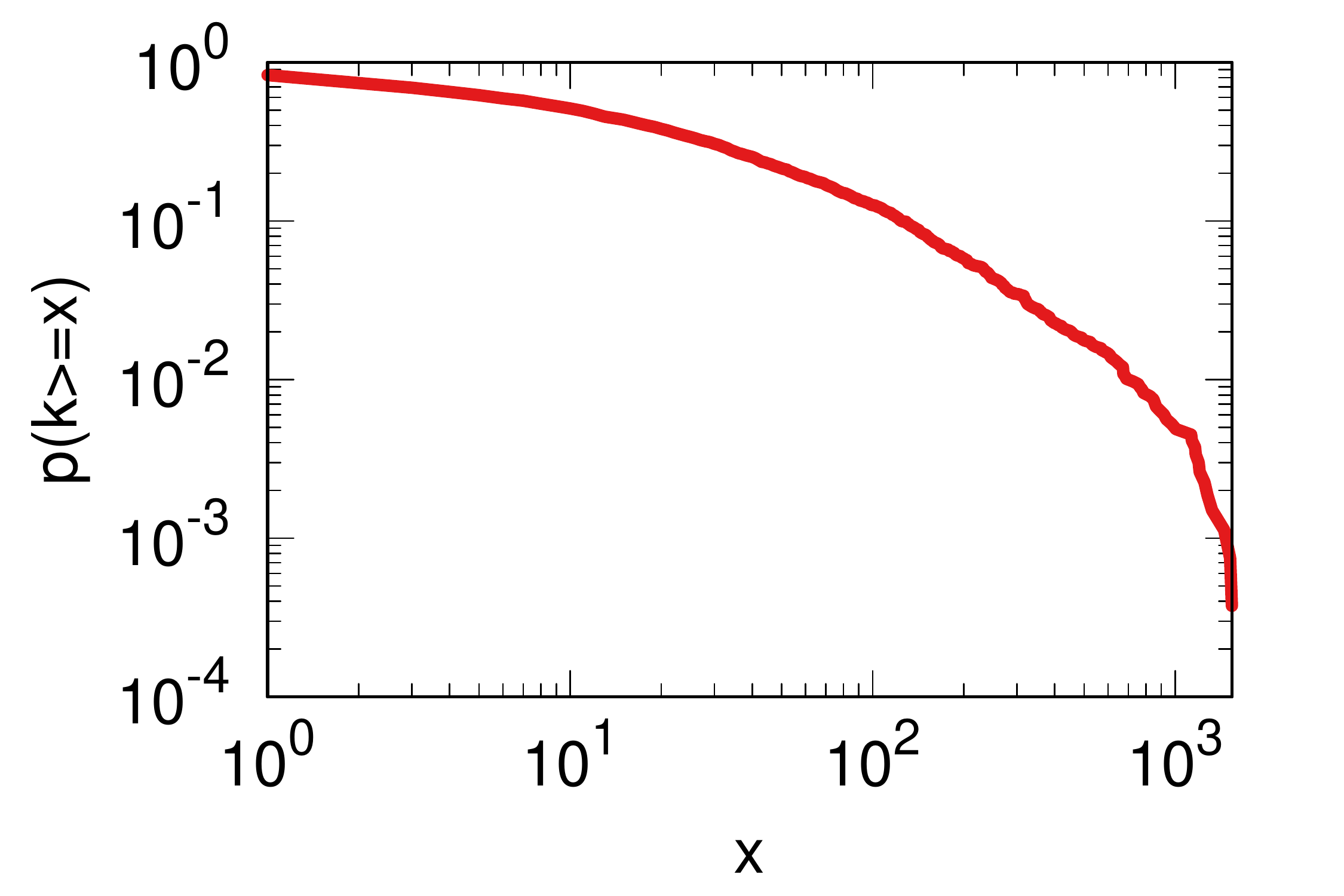}
\caption{The (complement of the) cumulative edge weight distribution of the full $ASN$: the probability (y-axis, log scale) that an edge has a weight equal to or larger than a certain value (x-axis, log scale).}
\label{fig:asn-ccdf}
\end{figure}

The quality measure is the codelength necessary to encode random walks. The codelength gives us a corroboration of the presence of communities. Without communities, we need $\sim 8.52$ bits to encode the random walks. With communities, the codelength reduces to $\sim 4.48$.

Figure \ref{fig:asn-ccdf} shows the complement of the cumulative distribution (CCDF) of the edge weights of $ASN$ before operating the backboning. We can see that, while the distribution is not a power-law -- note the log-log scale --, it nevertheless spans multiple orders of magnitude, with a clear skewed distribution. In fact, 50\% of the edges have a weight lower than 10 -- only in 10 cases out of the possible 960 + 819 the two algorithms were in the top five most similar results --, while the three strongest edges (.1\% of the network) have weights of 1,453, 1,519, and 1,540, respectively.

This means that the distribution could have been a power-law, had we performed enough tests. In any case, such broad distribution justifies our choice of backboning method, which is specifically designed to handle cases with large variance and lack of well-defined averages.

\subsection{Robustness}\label{sec:anal-robust}
In developing our framework, we made choices that have repercussions $ASN$'s shape. How much do these choices impact the final result? We are interested in estimating the amount of change in $ASN$'s topology, specifically whether it is stable: different $ASN$s calculated with different procedures and parameters are similar.

\begin{figure}
\centering
\includegraphics[width=.32\columnwidth]{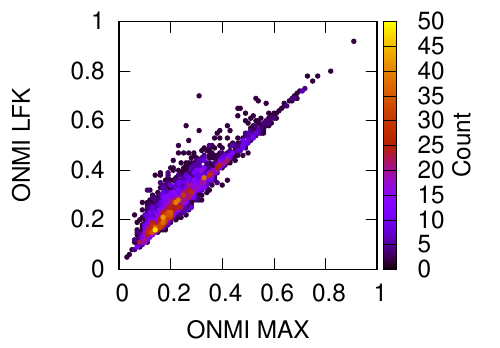}
\includegraphics[width=.32\columnwidth]{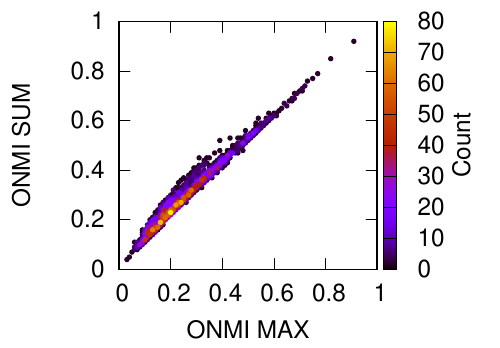}
\includegraphics[width=.32\columnwidth]{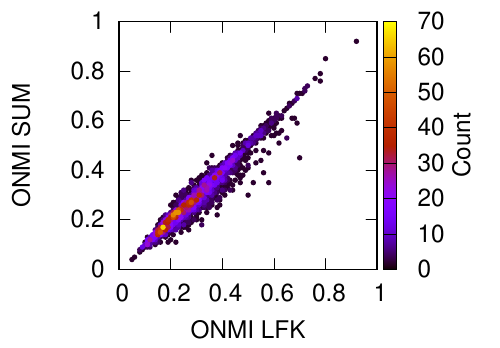}
\caption{The correlation between the $ASN$ weights using different oNMI variants: (left) MAX vs LFK; (middle) MAX vs SUM; (right) LFK vs SUM. Each dot is an algorithm pair and the color represent how many pairs shared a given oNMI score combination.}
\label{fig:robust-1}
\end{figure}

The first test aims at quantifying the amount of change introduced by using a different oNMI measure. Recall that our official $ASN$ uses the MAX variant. There are two alternatives: LFK and SUM. Figure \ref{fig:robust-1} shows how $ASN$s calculated using them correlated with the MAX standard version.

It is immediately obvious from the plots that the choice of the specific measure of oNMI has no effect on the shape of $ASN$. We could have picked any variant and we would have likely observed similar results. In fact, the correlations between the methods are as follows: MAX vs LFK = 0.94; MAX vs SUM = 0.99; LFK vs SUM = 0.97.

\begin{figure}
\centering
\includegraphics[width=.32\columnwidth]{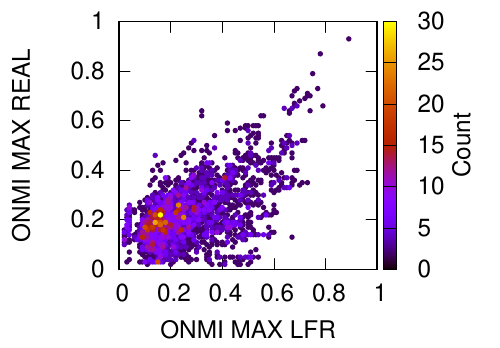}
\includegraphics[width=.32\columnwidth]{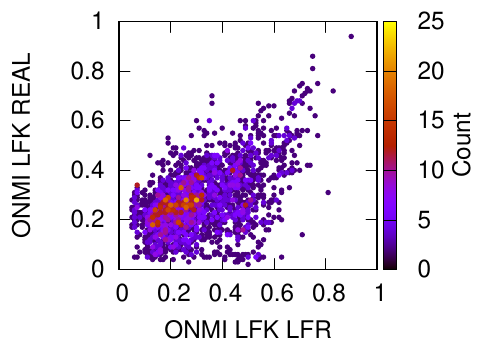}
\includegraphics[width=.32\columnwidth]{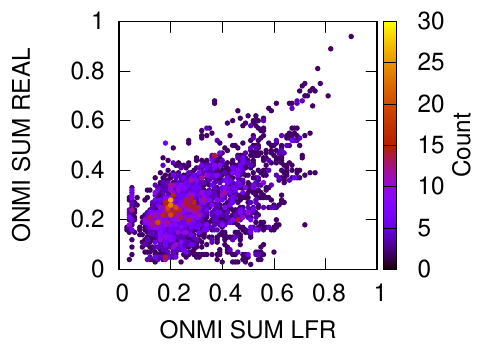}
\caption{Correlation between the $ASN$ weights using the LFR benchmarks (x-axis) and the real world networks (y-axis). Same legend as Figure \ref{fig:robust-1}, for different oNMI variants: (left) MAX, (middle) LFK, (right) SUM.}
\label{fig:robust-2}
\end{figure}

The second test focuses on the synthetic LFR benchmarks versus the 819 real world networks. Real world networks do not necessarily look like LFR benchmarks -- or each other. On the other hand, all LFR benchmarks are similar to each other. Does that create different $ASN$s? We repeat our correlation test (Figure \ref{fig:robust-2}). As in the previous cases, we observe a significant positive correlation for all tests -- albeit lower than before: LFR vs Real (MAX) = 0.55; LFR vs Real (LFK) = 0.51; LFR vs Real (SUM) = 0.51.

All these correlations are still statistically significant ($p \sim 0$). However, we concede that there is a difference between real world networks and LFR benchmarks. It is worthwhile investigating this difference in future works, as a possible argument against the blind acceptance of LFR as the sole benchmark for testing community discovery algorithms.

Third, our edge weights are a count of benchmarks in which two algorithms were in each other most similar lists. Alternative edge creation procedures might be to take the average oNMI, or to count the similarity between two algorithms only if they exceed a fixed oNMI threshold.

Section \ref{sec:method-simil} provides our theoretical reasons. Here we show that, at a practical level, our results are not gravely affected by such choice. We do so by calculating the NMI between $ASN$'s communities obtained with all three techniques. The $ASN$ built by averaging the similarity scores has a 0.63 NMI with our option, while the one obtained by a fixed threshold has a 0.46 NMI. On the basis of these similarities, we conclude that there is an underlying $ASN$ structure, and we think our choices allow us to capture it best.

\subsection{Communities}
In Figure \ref{fig:asn}, we show a partition of $ASN$ into communities. A seasoned researcher in the community discovery field would be able to give meaningful labels to those communities. Here, we objectively quantify this meaningfulness along a few dimensions of the many possible.

We start by considering a few attributes of community detection algorithms, whether they: return overlapping partitions (in which communities can share nodes), are based on some centrality measure (be it random walks or shortest paths) or spreading process (it will become apparent why we lump these two categories), are based on modularity maximization \cite{newman2006modularity}, or are based on a neighborhood similarity approach (e.g. they cluster the adjacency matrix).

\begin{table}
\centering
\begin{tabular}{ll|rrrrr}
ID & Col & $n$ & Over & Spr & Q & NSim\\
\hline
1 & Red & 21 & 0.9048 & 0.1429 & 0.0952 & 0.0952\\
2 & Blue & 28 & 0.3214 & 0.5357 & 0.1429 & 0.0357\\
3 & Green & 10 & 0.1000 & 0.0000 & 1.0000 & 0.0000\\
4 & Purple & 11 & 0.0909 & 0.0000 & 0.0000 & 0.7273\\
5 & Orange & 8 & 0.3750 & 0.2500 & 0.3750 & 0.0000\\
\end{tabular}
\caption{Features of the communities of $ASN$. $n$: \# of nodes. Over: \% overlapping algorithms. Spr: \% algorithms based either on centrality measures (including edge betweenness and random walks) or some sort of spreading process (e.g. label percolation). Q: \% algorithms based on modularity maximization. NSim: \% algorithms based on neighborhood similarity. Algorithms can be part of multiple/no classes, so the rows do not sum to one.}
\label{tab:comm-features}
\end{table}

In Table \ref{tab:comm-features} we calculate the fraction of nodes in a community in each of those categories. Note that we count overlap nodes in all of their communities, so some nodes contribute to up to three communities. As we expect, some communities have a stronger presence of a single category.

\begin{table*}
\centering
\begin{tabular}{ll|rrrrrr}
ID & Col & $\bar{|C|}$ & Avg Size & $\bar{d}$ & $\bar{Q}$ & $\bar{c}$ & Avg Ncut\\
\hline
1 & Red & 19.7979 & 9.0942 & 0.3220 & 0.2200 & 0.7423 & 0.7674\\
2 & Blue & 5.6520 & 16.4769 & 0.2627 & 0.1102 & 0.5542 & 0.7100\\
3 & Green & 4.8948 & 11.9844 & 0.2580 & 0.1118 & 0.6288 & 0.7407\\
4 & Purple & 10.3702 & 11.0140 & 0.2917 & 0.0333 & 0.7555 & 0.8033\\
5 & Orange & 4.2852 & 17.0505 & 0.2329 & 0.0863 & 0.5963 & 0.7483\\
\end{tabular}
\caption{The averages of various community descriptive statistics per algorithm group. $\bar{|C|}$: Average number of communities. Avg Size: Average number of nodes in the communities. $\bar{d}$: Average community density. $\bar{Q}$: Average modularity -- when the algorithm is overlapping we use the overlapping modularity instead of the regular definition. $\bar{c}$: Average conductance -- from \cite{leskovec2010empirical}. Avg Ncut: Average normalized cut -- from \cite{leskovec2010empirical}.}
\label{tab:comm-stats}
\end{table*}

The largest community (in blue) groups centrality-based algorithms (Infomap \cite{rosvall2008maps}, Edge betweenness \cite{newman2004finding}, Walktrap \cite{pons2005computing}, etc) with the ones based on spreading processes (label percolation \cite{raghavan2007near}, SLPA \cite{cordasco2010community}, Ganxis \cite{xie2013labelrank}, etc). Some of these can be overlapping, but the majority of nodes in the community is part of this ``spreading'' category. This community shows a strong relationship between random walks, centrality-based approaches, and approaches founded on spreading processes.

The second largest community (in red) is mostly populated by overlapping approaches (more than 90\% of its nodes are overlapping) -- BigClam \cite{yang2013overlapping}, k-Clique \cite{palla2005uncovering}, and DEMON \cite{coscia2012demon} are some examples. The third largest community (in purple) is mostly composed by algorithms driven by neighbor similarity (more than 70\% of them) rather than the classical ``internal density'' definition (the two are not necessarily the same). The fourth largest community (in green) exclusively groups modularity maximization algorithms.

We now calculate descriptive statistics of the groupings each method returns and then we calculate its average across all the test networks. To facilitate interpretation, we also aggregate at the level of the $ASN$ community, as we show in Figure \ref{fig:asn}. Table \ref{tab:comm-stats} reports those statistics.  We also calculate the standard errors, which prove that these differences are significant, but we omit them to reduce clutter.

The results from Table \ref{tab:comm-stats} can be combined from the knowledge we gathered from Table \ref{tab:comm-features}. For instance, consider community 4. We know from Table \ref{tab:comm-features} that this hosts peculiar algorithms working on ``neighbor similarity'' rather than internal density. This might seem like a small difference, but Table \ref{tab:comm-stats} shows its significant repercussions: the average modularity we get from these algorithms is practically zero. Moreover, the algorithms tend to return more -- and therefore smaller -- communities, which tend to be denser but also to have higher conductance.\footnote{Community 1 returns more communities, but it is composed by overlapping algorithms, which can return more communities without necessarily make them small, as they can share nodes. Thus its communities are larger than one would expect given their number.} This is another warning sign for uncritically accepting modularity as the \textit{de facto} quality measure to look at when evaluating the performance of a community discovery algorithm. It works perfectly for the methods based on the same community definition, but there are other -- different and valid -- community definitions.

Other interesting facts include the almost identical average modularity between community 2 -- whose algorithms are explicitly maximizing modularity -- and community 3 -- which is based on spreading processes. Community 1 has higher internal density, but also higher conductance and normalized cut than average, showing how overlapping approaches can find unusually dense communities, sacrificing the requirement of having few outgoing connections.

\begin{figure}
\centering
\includegraphics[width=\columnwidth]{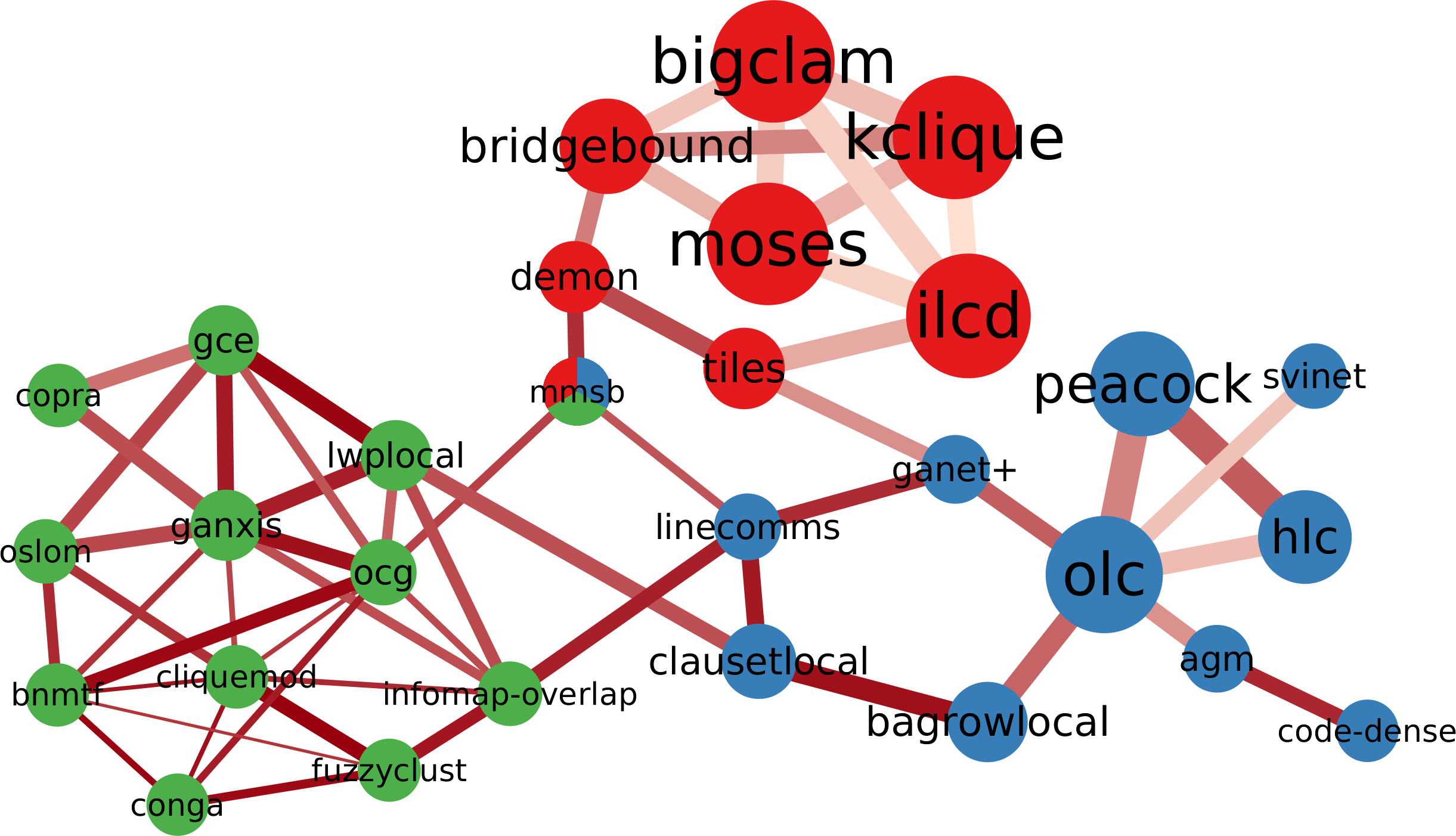}
\caption{The $ASN$ focusing exclusively on overlapping community discovery algorithms. The legend of the figure is the same as the one for Figure \ref{fig:asn}.}
\label{fig:asn-overlap}
\end{figure}

The categories we discussed are necessarily broad and might group algorithms that have significant differences in other aspects. For instance, there are hundreds of different ways to make your algorithm return overlapping communities -- communities sharing nodes. Our approach allows us to focus on such methods to find differences inside the algorithm communities. In practice, we can generate different versions of $ASN$, by only considering the similarities between the algorithms in the ``overlapping'' category.

Note that this is different than simply inducing the graph from the original $ASN$, selecting only the overlapping algorithms and all the edges between them. Here we select the nodes and all their similarities and then we apply the backboning, with a different -- higher -- $\delta$ threshold. In this way, we can deploy a more stringent similarity test, that is able to distinguish between subcategories of the main category.

Figure \ref{fig:asn-overlap} depicts the result. Infomap divides the overlapping $ASN$ in three communities, proving the point that there are substantial sub-classes in the overlapping coverage category. There are strong arguments in favor of these classes being meaningful, although a full discussion requires more space and data. For instance, consider the bottom-right community of the network (in blue). It contains all the methods which apply the same strategy to find overlapping communities: rather than clustering nodes, they cluster edges. This is true for Linecomms \cite{evans2009line}, HLC \cite{ahn2010link}, Ganet+ \cite{pizzuti2009overlapped}, and OLC \cite{ball2011efficient}. The remaining methods do not cluster link directly, but $ASN$ suggests that their strategies might be comparable.

We can conclude that $ASN$ provides a way to narrow down to subcategories of community discovery and find relevant information to motivate one's choice of an algorithm.

\subsection{Ground Truth in Synthetic Networks}
The version of $ASN$ based on synthetic LFR benchmarks allows an additional analysis. The LFR benchmark generates a network with a known ground truth: it establishes edges according to a planted partition, which it also provides as an output. Thus, we can add a node to the network: the ground truth. We calculate the similarity of the ground truth division in communities with the one provided by each algorithm. We now can evaluate how the algorithms performed, by looking at the edge weights between the ground truth node and the algorithm itself. In the MAX measure, this means the number of times the algorithm was in the top similarity with the ground truth and vice versa.

\begin{table}
\centering
\begin{tabular}{l|l|r}
Rank & Algorithm & oNMI MAX\\
\hline
1 & linecomms & 165\\
2 & oslom & 73\\
3 & infomap-overlap & 64\\
4 & savi & 62\\
5 & labelperc & 57\\
6 & rmcl & 54\\
7 & edgebetween & 41\\
7 & leadeig & 41\\
7 & vbmod & 41\\
10 & gce & 32\\
\end{tabular}
\caption{The ten nodes with the highest MAX edge weight with the ground truth node in $ASN$ -- using exclusively data from the LFR synthetic networks.}
\label{tab:rankings}
\end{table}

Table \ref{tab:rankings} shows the ten best algorithms in our sample. We do not show the worst algorithms, because MAX is a strict test, and thus there is a long list of (21) algorithms with weight equal to zero, which is not informative. The table shows that the best performing algorithm are Linecomms, OSLOM, and the overlap version of Infomap.

Should we conclude that these are the best community discovery algorithms in the literature? The answer is yes only if we limit ourselves to the task of finding the same type of communities that the LFR benchmark plants in its output network. Crucially, this does not include all possible types of communities you can find in complex networks. To see why this is the case, consider again $ASN$ from Figure \ref{fig:asn}. The ten nodes listed in Table \ref{tab:rankings} are not scattered randomly in the network: they tend to be in the same area. Specifically we know that the ground truth node is located deep inside the blue community, as most of the top ten algorithms from Table \ref{tab:rankings} are classified in that group.

\begin{figure}
\centering
\includegraphics[width=.66\columnwidth]{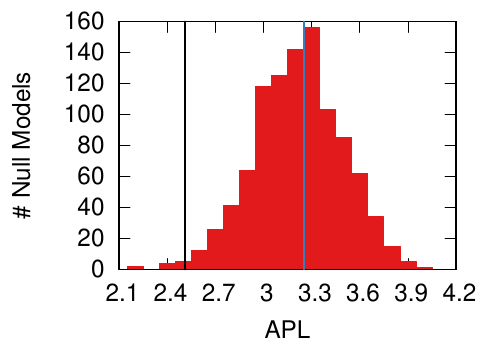}
\caption{The number of sets of 10 random nodes (y-axis) with a given avg path length between them. The black line shows the observation. The blue line shows the average path length of $ASN$.}
\label{fig:apl-null-expectation}
\end{figure}

We can quantify this objectively by calculating the average path length between the ten nodes, which is equal to 2.51 -- on average you need to cross two and a half edges to go from any of these ten nodes to any other of the ten. This is shorter than the overall average path length in $ASN$, which is 3.25. We test statistical significance by calculating the expected average path length when selecting ten random nodes in the network. Figure \ref{fig:apl-null-expectation} shows the distribution of their distances. Only seven out of a thousand attempts generated a smaller or equal average path length.

We conclude this section with a word of caution when using benchmarks to establish the quality of a community discovery algorithm, which is routinely done in review works and when proposing a new approach. If the benchmark does not fit the desired definition of community, it might not return a fair evaluation. If one is interested in communities based on neighborhood similarity -- the green community in Figure \ref{fig:asn} -- the LFR benchmark is not the correct one to use. Moreover, when deciding to test a new method against the state of the art, one must choose the algorithms in the literature fitting the same community definition, or the benchmark test would be pointless. This warning goes the other way: assuming that all valid communities look like the ones generated by the LFR benchmark would impoverish a field that -- as the strong clusters in $ASN$ show -- does indeed have significantly different perspectives of what a community is.

\section{Conclusion}
In this paper we contributed to the literature on reviewing community discovery algorithms. Rather than classify them by their process, community definition, or performance, here we classify them by their similarity. How similar are the groupings they return? We performed the most comprehensive analysis of community discovery algorithms to date, including 73 algorithms tested over more than a thousand synthetic and real world networks. We were able to reconstruct an Algorithm Similarity Network -- $ASN$ -- connecting algorithms to each other based on their output similarity. $ASN$ confirms the intuition about the community discovery literature: there are indeed different valid definitions of community, as the strong clustering in the network shows. The clusters are meaningful as they reflect real differences among the algorithms' features. $ASN$ allows us to perform multi-level analysis: by focusing on a specific category, we can apply our framework to discover meaningful sub-categories. Finally, $ASN$'s topology highlights how projecting the community detection problem on a single definition of community -- e.g. ``a group of nodes densely connected to each other and sparsely connected with the rest of the network'' -- does the entire sub-field a disservice, by trivializing a much more diverse set of valid community definitions.

By its very nature, this paper will always be a work in progress. We do not claim that there are only 73 algorithms in the community discovery literature that are worth investigating. We only gathered what we could. Future work based on this paper can and will include whatever additions authors in the field feel should be considered -- and they are encouraged to help us by sending suggestions and/or working implementations to \url{mcos@itu.dk}. The most up to date version of $ASN$ will be available at \url{http://www.michelecoscia.com/?page_id=1640}. Moreover, for simplicity, here we focused only on algorithms that work on the simplest graph representations. Several algorithms specialize in directed, multilayer, bipartite, and/or metadata-rich graphs. These will be included as we refine the $ASN$ building procedure in the future.

\bibliographystyle{ACM-Reference-Format}

\bibliography{biblio}

\end{document}